\documentclass[useAMS,usenatbib]{mn2e}

\usepackage[T1]{fontenc}
\usepackage{aecompl}

\usepackage{times,graphicx,amssymb,natbib,subfigure}

\newcommand{\mjup}{M_{\rm Jup}}
\newcommand{\msol}{M_{\rm \odot}}

\title[Dynamics of Tidally Downsized Disc Fragments]{The Dynamical Fate of Self-Gravitating Disc Fragments After Tidal Downsizing}
\author[Duncan Forgan, Richard J. Parker and Ken Rice]{Duncan Forgan $^{1,2}$\thanks{E-mail:
dhf3@st-andrews.ac.uk}, Richard J. Parker$^{3}$ and Ken Rice$^{1}$ \\
$^{1}$Scottish Universities Physics Alliance (SUPA), Institute for Astronomy, University of Edinburgh, Blackford Hill, Edinburgh, EH9 3HJ, Scotland, UK \\
$^{2}$Scottish Universites Physics Alliance (SUPA), School of Physics and Astronomy, University of St Andrews, North Haugh, St Andrews, KY16 9SS \\
$^{3}$Astrophysics Research Institute, Liverpool John Moores University, 146 Brownlow Hill, Liverpool, L3 5RF, UK }

\begin{document}

\date{Accepted}

\pagerange{\pageref{firstpage}--\pageref{lastpage}} \pubyear{}

\maketitle

\label{firstpage}

\begin{abstract}

\noindent The gravitational instability model of planet/brown dwarf formation proposes that protostellar discs can fragment into objects with masses above a few Jupiter masses at large semimajor axis. Tidal downsizing may reduce both the object mass and semimajor axis. However, most studies of tidal downsizing end when the protostellar disc disperses, while the system is embedded in its parent star-forming region. To compare disc fragment descendants with exoplanet and brown dwarf observations, the subsequent dynamical evolution must be explored.

We carry out N -Body integrations of fragment-fragment scattering in multi-object star systems, and star systems embedded in substructured clusters. In both cases, we use initial conditions generated by population synthesis models of tidal downsizing.
The scattering simulations produce a wide range of eccentricities. The ejection rate is around 25\%. The ejecta mass distribution is similar to that for all objects, with a velocity dispersion consistent with those produced by full hydrodynamic simulations.  The semimajor axis distribution after scattering extends to parsec scales.

In the cluster simulations, 13\% of objects are ejected from their planetary system, and around 10\% experience significant orbit modification. A small number of objects are recaptured on high eccentricity, high inclination orbits. The velocity distribution of ejecta is similar to that produced by fragment-fragment scattering.

If fragment-fragment scattering and cluster stripping act together, then disc fragmentation should be efficient at producing free-floating substellar objects, and hence characterising the free-floating planet population will provide strong constraints on the frequency of disc fragmentation.

\end{abstract}


\section{Introduction}

\noindent Disc fragmentation through gravitational instability has been considered a potential formation channel for giant planets and very low mass stars for some time \citep{Cameron1978,Boss_science}.  Self-gravitating protostellar discs that are prone to  fragmentation produce objects with masses greater than a few Jupiter masses \citep{Forgan2011a} at semimajor axes larger than $\sim$ 30 AU \citep{Rafikov_05, Matzner_Levin_05, Whit_Stam_06,Mejia_3,Stamatellos2008, intro_hybrid, Clarke_09,Kratter2009,Vorobyov2010,Forgan2011a, Forgan2013, Tsukamoto2014}.  Once the conditions for fragmentation are met, the process occurs rapidly, typically on dynamical timescales.  Simulations of collapsing molecular clouds show that when discs fragment, it is typically within one free-fall time of the star's own birth \citep{Walch2009,Walch2010,collapses}, which is somewhat necessary given that self-gravitating discs produce strong, non-local angular momentum transport that can quickly drain the disc's mass and curtail fragmentation \citep{Harsono2011, Forgan2011}.

While much of the early work in disc instability theory focused on the fragmentation process, recent efforts have attempted to establish the final products of disc fragmentation - for example, the recently-coined tidal downsizing hypothesis \citep{Nayakshin2010b,Nayakshin2010,Nayakshin2010a} which builds on earlier work on fragment evolution \citep{Helled2008,Helled2008a,Boley2010b,Boley2011}.  Each fragment (hereafter embryo) contains a sample of the disc's population of dust grains.  It can add to this sample by accreting from the disc \citep{Helled2006} and grain size evolution can continue.  Grains will grow and sediment to the pressure maximum at the centre of the embryo, and can potentially form a core if the solids' density is sufficiently high.  

During this process, the embryo's position in the disc is also changing. Migration due to disc torques can move the embryo inward, and can result in tidal stripping if the embryo's contraction rate is slow.  In extreme cases, this process can completely disrupt the embryo, bequeathing its evolved grain populations back to the disc \citep{Nayakshin2012} and potentially causing outburst events as the embryo's remains are accreted by the star \citep{Dunham2012,Nayakshin2012a, Nayakshin2013}.

Embryos can escape some of these dangers by undergoing a final collapse to a bound object, either due to the dissociation of hydrogen molecules as in the case of young protostars (cf \citealt{Masu_98}), or due to hydrostatic instability in the embryo's gaseous, metal-rich envelope around its newly forming core \citep{Nayakshin2014a}.  This may allow them to survive even at relatively low semimajor axes.  

Recent population synthesis models \citep{TD_synthesis, Galvagni2013}, taking some or most of these effects into account, have extended the modelling timescales of fragmentation from a few tens of thousands of years up to around 1 Myr.  However, these timescales are still too short compared with observations of potential disc fragmentation descendants.  For example, the planetary system HR 8799 is often regarded as a candidate for disc fragmentation \citep{Nero2009}, with at least four $\sim 10 \mjup$ planets orbiting at 15-70 au from the host star.  However, the minimum age of the system is approximately 30 Myr \citep{Baines2012}.  If disc fragmentation modelling typically ends at the epoch of disc dispersal ($\sim$ 1 Myr) then how can we be confident that the systems produced remain in the same configuration? 

The low eccentricities observed in the HR 8799 system appears to be an example of a system stabilised by a mean motion resonance \citep{Godziewski2014} but it is possible that disc fragments can be excited to high eccentricities.  Systems with multiple orbiting objects are likely to experience dynamical instabilities if the orbital separations become sufficiently low \citep{Chambers1996,Papaloizou2001,Chatterjee2008}.  Such systems allow planet pairs to undergo close encounters, which can result in planet-planet scattering.  These events excite orbital eccentricities, after which tidal circularisation could be responsible for short period planets such as Hot Jupiters \citep{Weidenschilling1996,Fabrycky2007}, as well as the ``tightly packed'' multiple planetary systems revealed by the Kepler Space Telescope \citep{Raymond2009a}.  In some cases, the eccentricity is excited beyond unity, ejecting one or more objects from the system \citep{Rasio1996}.   Planet-planet scattering has been invoked for core accretion theory to explain the eccentricity distribution in current exoplanet data \citep{Adams2003,Ford_and_Rasio_08, Juric2008}.  Given that multiple disc fragments can form in a single system, it is not unreasonable to expect that fragment-fragment scattering to also be important, even while disc gas is still present \citep[cf][]{Moeckel2008}.

We should therefore expect ejected disc fragments to be progenitors of a variety of substellar objects.  In particular, we should expect to see signatures in the free-floating planet population \citep{Sumi2011}.  The origin of these objects remains the subject of debate.  Planet-planet scattering appears to be insufficient to explain the entire free floating population \citep{Veras2012}, but it presumably plays an important role in shaping it. \emph{In situ} formation of substellar objects in molecular clouds is possible for objects of tens of Jupiter masses \citep{Hennebelle2008,Hennebelle2009,Strigari2012,Andre2012, Palau2014}, and tidal stripping of substellar companions while inside the parent cluster is likely to play some role \citep{Parker2012,Craig2013}\footnote{We should also note that stellar mass loss in the post-main sequence epoch could produce free-floating planets \citep{Mustill2013,Veras2013}, but we shall focus on younger systems in this work.}.  

Forming more massive objects at larger semimajor axis increases the likelihood of stripping by the host cluster \citep{Boley2012}, and scattering events can be more energetic with more massive objects (although potentially less frequent due to larger orbital periods).  This would suggest that disc fragmentation may be extremely efficient at forming free floating planetary mass objects and field brown dwarfs.  

In fact, \emph{if observations characterise the free-floating planet population to a sufficient level of detail and accuracy, then this will place important constraints on the fraction of protostellar discs that fragment}, a hitherto unknown quantity that will be essential for disc fragmentation theory to have any predictive power.

The current mass distribution of free-floating objects does appear to have a sharp cutoff near the theoretical prediction of minimum initial fragment mass, viz. a few Jupiter masses \citep{Sumi2011}, but this is probably due to observational limitations.  We should also note that re-analysis of the sample gives reduced numbers of objects classified as unbound \citep{Quanz2012}.  

In any case, we should make predictions for future microlensing missions such as WFIRST \citep{Spergel2013}, which will be able to place tighter constraints on the free floating planet population.  Equally, the bound, large semimajor axis population of planets and brown dwarf companions will also be probed by WFIRST, providing further constraints on the frequency of disc fragmentation in the Milky Way.

In this work we attempt to ascertain the dynamical fate of disc fragments after the protostellar disc has dispersed.  We use data from \citet{TD_synthesis}'s population synthesis models of disc fragmentation as input to two separate studies of $N$-Body dynamics.  In the first, we consider isolated planetary systems, and carry out fragment-fragment scattering experiments to ascertain the subsequent distribution of orbital elements and ejection rates.  In the second, we place planetary systems within clusters to investigate orbital evolution and tidal stripping.  In section \ref{sec:methods} we briefly describe the population synthesis model and the $N$-Body experiments; in section \ref{sec:results} we describe the results, in section \ref{sec:discussion} we discuss the implications for exoplanet and free floating planet observations, and in section \ref{sec:conclusions} we summarise the work.

%

\section{Methods} \label{sec:methods}

\subsection{The Initial Dataset}

\noindent We take our sample of planetary systems formed by tidal downsizing from the population synthesis models of \citet{TD_synthesis}.   We summarise this model briefly here.

We generate a large number of protostellar disc systems, which are evolved using a standard $\alpha$-viscosity \citep{Shakura_Sunyaev_73, Rice_and_Armitage_09, Clarke_09}, where $\alpha$ represents a turbulent viscosity produced by self-gravity.  The disc masses are selected to ensure that this pseudo-viscous approximation is not invalidated by non-local angular momentum transport \citep{Lodato2005,Forgan2011}.  The transport of angular momentum due to self-gravity results in accretion of disc material onto the central star. The discs are also subject to mass loss by X-Ray photoevaporation driving winds \citep{Owen2010,Owen2011}.  The disc-to-star mass ratio is fixed, but the central star mass and the X-Ray luminosity are independently varied, producing a variety of disc lifetimes consistent with observations \citep{Haisch2001}.  

In each of these protostellar discs, we determine sites of disc fragmentation by demanding that the Jeans mass inside a spiral arm perturbation be decreasing rapidly.  By calculating both the Jeans mass and its rate of change, we determine where a disc will fragment, and also the initial mass of the fragment \citep{Forgan2011,Forgan2013}.  The fragments are initially on circular orbits, and are separated by 1.5 to 3 Hill Radii.  The discs possess anywhere between 1 and 5 fragments.

The fragments are then evolved according to the semi-analytic framework of Nayakshin \citep{Nayakshin2010a,Nayakshin2010b,Nayakshin2010}.  The initial stages of fragment collapse are akin to those of the first cores (FCs) of the star formation process \citep{Masu_98}.  Grains grow and sediment once they reach a critical size to become sensitive to the local gas drag.  This sedimentation is suppressed by turbulence and by high-velocity collisions shattering grains.  Once the density of grains in the centre of the fragment exceeds the local gas density, the grains become self-gravitating and rapidly establish a core.  As the fragment continues to collapse, the central temperature rises.  Once this temperature exceeds $\sim$ 1600K, core formation is halted due to grain vapourisation.  Note that we do not include effects such as the core assisted gas capture instability recently described by \citet{Nayakshin2014a}.

During the evolution of the fragment and its solids population, it also migrates in the disc according to standard Type I/Type II migration timescales.  If the fragment radius exceeds its Hill Radius, then the upper layers of the fragment will be tidally stripped, and the mass of the fragment decreases.  While all fragments are initially several Jupiter masses, and hence firmly in the Type II regime, tidal stripping can move them to a Type I migration regime, causing the fragment's inward radial velocity to rapidly increase, and potentially completely destroying the fragment as the Hill Radius continues to shrink.  Fragment destruction is a common outcome from the population synthesis model - typically around 40\% of all fragments suffer this fate.  Note that this may be an underestimate of the true destruction rate, as more sophisticated migration modelling in self-gravitating discs suggest that migration timescales are shorter than the standard ones used here \citep{Baruteau2011}.

\begin{figure}
\begin{center}
\includegraphics[scale=0.4]{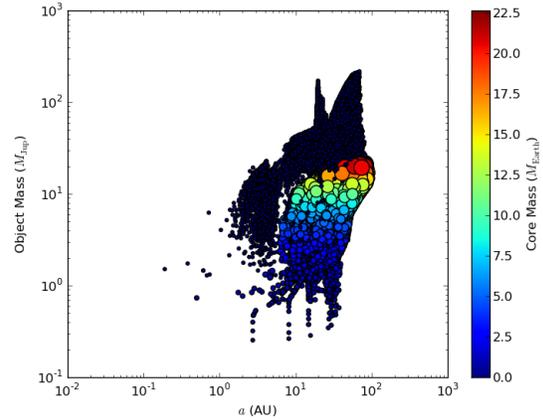}
\caption{The mass-semimajor axis distribution retrieved from the population synthesis model of \citet{TD_synthesis}.  This dataset includes brown dwarfs (objects with masses greater than 13 $\mjup$), and giant planets with and without cores. \label{fig:ics}}
\end{center}
\end{figure}

By repeating this process many times, we generate a population of objects.  In \cite{TD_synthesis} we ran four different iterations of the model to create four separate populations, investigating the effects of modifying various free parameters.  In this work, we use outputs from a single run of the model, corresponding to an opacity powerlaw index of 1, and ``standard'' migration (i.e., the migration timescale is not arbitrarily increased or decreased).  The initial distribution of object mass and semi-major axis can be found in Figure \ref{fig:ics}.

\noindent

\subsection{Fragment-Fragment Scattering \label{sec:method_scattering}}


\noindent The population synthesis model generates individual systems with multiple objects, providing a large number of systems to integrate.  We sample 10,000 systems from the above dataset, with a total of 23,325 orbiting objects in the sample.  Each system has at least two bodies in orbit of the host star: 2-body star-planet systems are not investigated.  

The population synthesis model gives a star mass, and orbiting object masses and semi-major axes.  It assumes that all orbits are circular, and it gives no information as to the relative phases of the orbits.  Given our ignorance of these parameters, we assume that the eccentricity is initially zero, all bodies are coplanar, and we initially assign orbital mean anomalies uniformly between $[0,2\pi]$.  The $N$-Body problem is solved using a 4th order Hermite integrator, with an adaptive global timestep.  The integrator conserves energy to 1 part in $10^6$ over the course of each simulation.  Note that we do not consider tidal evolution of the planet's orbit due to the star \citep[e.g.][]{Jackson2008}, and hence planets with highly eccentric orbits do not experience any eccentricity damping.  Collisions are not modelled in this analysis.

Each system is evolved for $10^6$ years.  This is still somewhat smaller than the observed ages of objects which we would like to compare to disc fragment models.  We select this relatively low simulation time to reduce computational expense and evolve more systems - typically, systems that produce scattering events express this instability within a few tens of thousands of years, which is to be expected given the objects' relatively close orbital spacing \citep{Chambers1996,Chatterjee2008}.  After this phase of instability, systems generally settle to quasistable orbital configurations.  Selecting this relatively short integration time will restrict our predictive power to some degree: for example, some systems will undergo eccentricity pumping that may lead to ejection on timescales greater than $10^6$ years.  In this sense we give conservative limits on both the eccentricity distribution and the ejection rate due to scattering.

\subsection{Simulating a Clustered Environment}{\label{sec:method:cluster}}

Most star-forming regions are observed to be filamentary \citep[e.g.][]{Arzoumanian2011}, which naturally leads to a hierarchically self-similar, or substructured spatial distribution of stars \citep{Cartwright2004,Sanchez2009,Gouliermis2014}. This substructure appears to be erased over time in observed star-forming regions \citep{Schmeja2008,Sanchez2009}, which is a consequence of 2-body relaxation \citep{Parker2012d,Parker2014b}.

\citet{Parker2014b} showed that dense regions that form with substructure and sub-virial or virial velocities will form a bound cluster \citep[see also][]{Allison2010}, whereas those with globally supervirial velocities (i.e.\,\,unbound) will retain substructure and possibly form large-scale association-like complexes. 

The key parameter for assessing the impact of the star-forming environment on planetary systems is the initial local density. \citet{Parker2012} show that even a supervirial (unbound) region which is dense will alter the orbits of planetary systems. 

It is important to note that a significant fraction of the dynamical interactions which will affect the planetary orbits occur in the early stages, even before the cluster has formed from the substructured star-forming region. For this reason, the effects of the star-forming environment will occur \emph{at the same time as the fragment-fragment scattering}. 

Currently, simulations that simultaneously evolve multi-planetary systems within a clustered environment are in their infancy due to technical limitations, although recent progress has been made \citep{Hao2013,Liu2013,Pacucci2013}. 

In this paper, we perform direct N-body star cluster simulations, in which the effects of interactions in the cluster on planetary-mass objects are explicitly computed (assuming the planet forms a binary system with its host star).  We randomly sample plantetary masses and semi-major axes from the distribution in \citet{TD_synthesis} before fragment--fragment scattering takes place, and determine the fraction that a) become free-floating, b) have their eccentricities significantly altered and c) have their semimajor axes significantly altered in the clustered environment. We then also use the simulations of \citet{Parker2012} with 1\,M$_{\rm Jup}$ single planets placed at 1, 5, 30 and 100\,au, which enables us to determine the fraction of systems that will be affected by the clustered environment which first have had their orbits altered by the fragment--fragment scattering experiments.

Our $N$-body simulations of star-forming regions are set-up in the same way as those in \citet{Parker2012}. The regions each have 750 stars drawn from  \citet{Kroupa2002} initial mass function of the form:
\begin{equation}
 N(M)   \propto  \left\{ \begin{array}{ll} 
 M^{-1.3} \hspace{0.4cm} m_1 < M/{\rm M_\odot} \leq m_2   \,, \\ 
 M^{-2.3} \hspace{0.4cm} m_2 < M/{\rm M_\odot} \leq m_3   \,,
\end{array} \right.
\end{equation}
and we choose $m_1$ = 0.1\,M$_\odot$, $m_2$ = 0.5\,M$_\odot$, and  $m_3$ =
50\,M$_\odot$.

If a star has a mass of $>$2.5\,M$_\odot$ we pair it in a binary system with another star drawn from a flat mass ratio distribution \citep{Metchev2009,Reggiani2011,Reggiani2013} and a semimajor axis and orbital eccentricity drawn from the period distribution of binaries in the field \citep{Raghavan2010}\footnote{The binary properties of massive stars may actually differ signficantly from the Solar-mass stars in \citet{Raghavan2010}. However, the presence (or not) of binaries in a cluster does not change the resulting fraction of planets that are affected by dynamics \citep{Parker2012}, so the exact details of their set-up do not matter.}.  This is to prevent massive stars that may evolve significantly over the 10\,Myr timescale of the simulations from being paired with a planet. In the first set of simulations, all stars with lower masses are paired with a planet with mass and semimajor axis from the distribution in \citet{TD_synthesis}\footnote{In reality, many stars $<$2.5\,M$_\odot$ are also in binary systems. However, in order to fully sample the distribution from \citet{TD_synthesis} we pair each star with a planet, rather than a stellar companion.}, and in the second set of simulations, the planets all have a mass of 1\,M$_{\rm Jup}$ and are placed at 1, 5, 30 and 100\,au from the host star. The planets have zero eccentricity initially.  

We note that this set-up may in fact over produce systems with planets formed via GI. However, the purpose of this paper is to perform a simple numerical experiment to assess the fractions of these planets that may then become free-floating due to interactions in a clustered environment. The initial stellar binary distribution that we adopt may not be an accurate representation of that in star-forming regions (which are largely unconstrained, e.g. \citealt{King2012}), however it is likely that the initial population is field-like \citep{Bate2012, Parker2014}. The deficit of stellar binaries in our calculations may underestimate the fraction of planets that become free-floating, as an disruptive interaction with a (relatively) massive binary system would be more likely than an interaction with a single star or star-planet system due to the higher collisional cross-section of that binary, although in earlier work \citet{Parker2012} do not find a strong dependence on stellar binarity.

The star-star and star-planet systems are then placed randomly in a substructured fractal distribution, as described in \citet{Goodwin2004a,Allison2010,Parker2014b}. A fractal distribution is the most convenient way of creating hierarchical substructure, because the level of substructure is quantified by a single number, the fractal dimension, $D$. Star-forming regions may, or may not be truly fractal \citep{Elmegreen2001}, but can usually be described by a fractal dimension \citep{Cartwright2004}.

Our fractal star-forming regions have a moderate amount of substructure $D = 2.0$, and a radius of 1\,pc. This results in a median density of $\sim$5000\,M$_\odot$\,pc$^{-3}$ in our simulations, which is somewhat higher than most local star-forming regions \citep{Bressert2010}. However, the `average' or dominant star-forming region which produces planet-hosting field stars is still unknown, and our simulations are consistent with the initial conditions which may have formed nearby star clusters such as the ONC \citep{Allison2010,Allison2011,Parker2011c}. For this reason, our simulations of the evolution of star-forming region probably represent an upper limit to the maximum density that planetary systems experience during their formation. The virial ratio of our star-forming regions is set to $\alpha_{\rm vir} = 0.3$, so that they are sub-virial and collapse to form a bound cluster after 1\,Myr.

The star-forming region are evolved for 10\,Myr using the \texttt{kira} integrator in the \texttt{Starlab} environment \citep[e.g.][]{Zwart1999,Zwart2001}, which utilises a 4$^{th}$-order Hermite scheme. We check whether a star--planet system is bound by determining the total energy of the system and the proximity of nearest neighbours \citep{Parker2009}. If the system has negative total energy, and the particles are mutual nearest neighbours, we consider it to be bound.  We do not include stellar evolution in the simulations. 

\section{Results}\label{sec:results}

\subsection{Fragment-Fragment Scattering}\label{sec:results:frag}

\subsubsection{Properties of Bound Objects}

\begin{figure*}
\begin{center}$
\begin{array}{cc}
\includegraphics[scale=0.4]{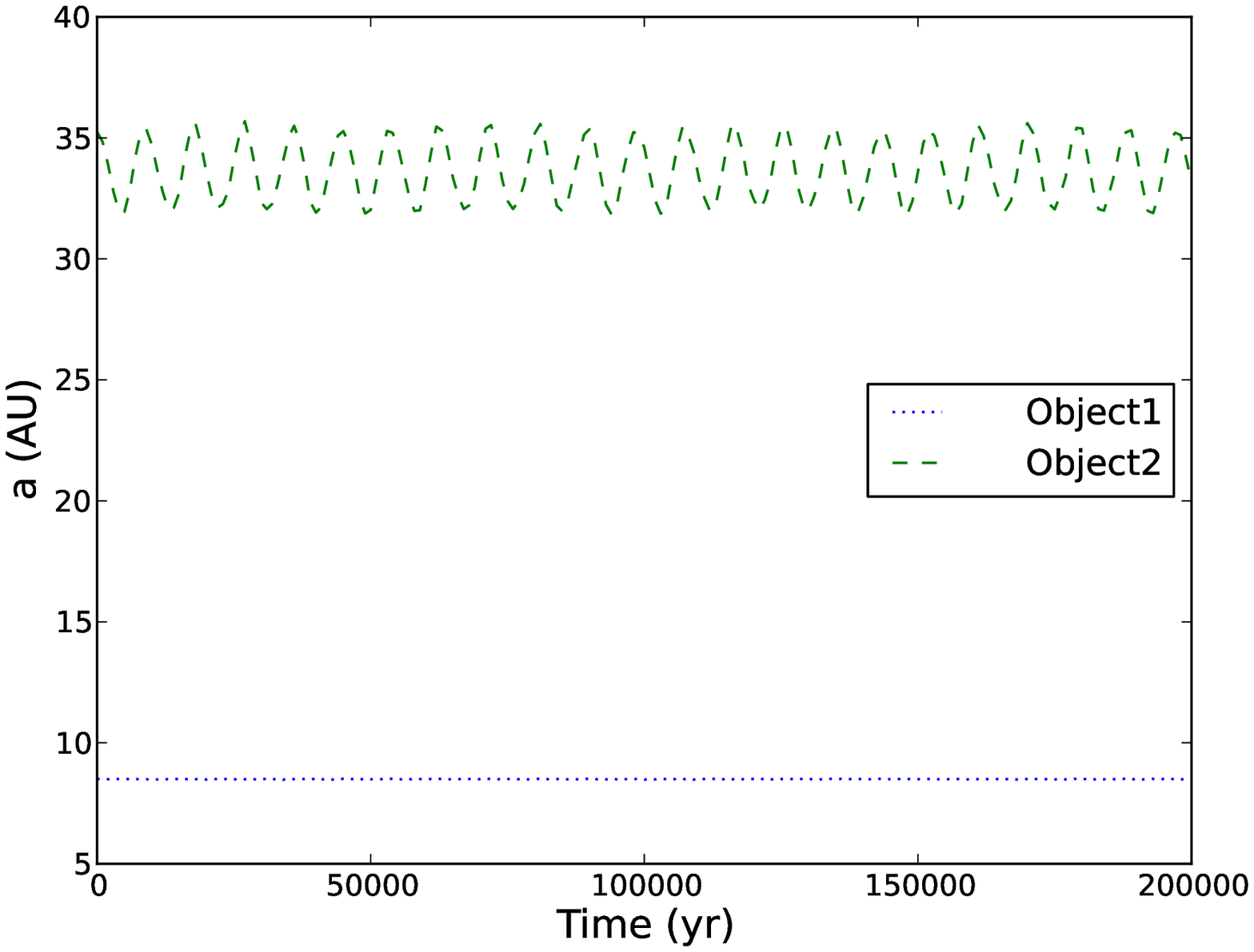} &
\includegraphics[scale=0.4]{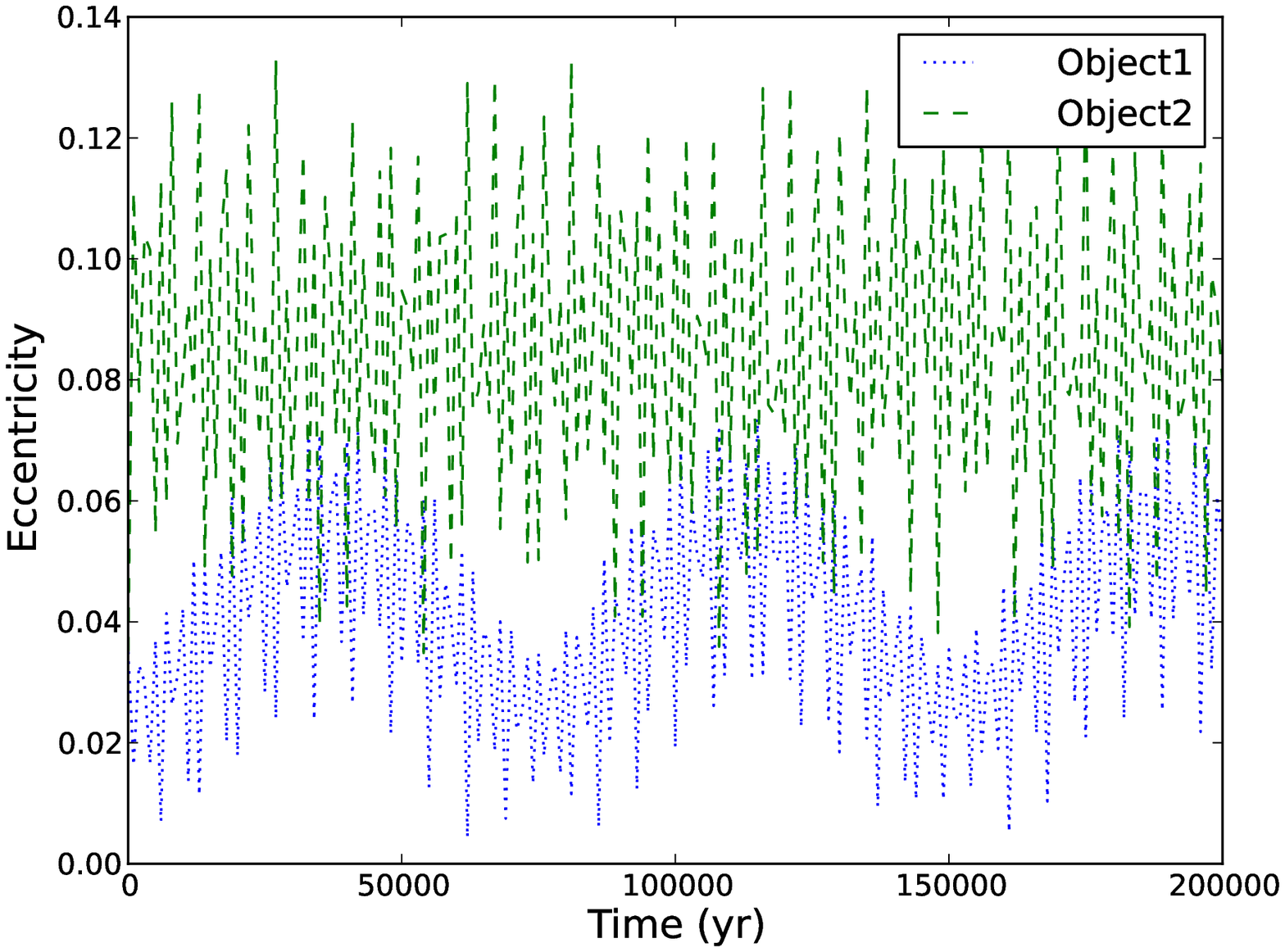} \\
\end{array}$
\caption{Semimajor axis (left) and eccentricity (right) as a function of time for a single system that maintains a stable configuration.  The central star mass is 1.18 $\msol$, and the 2 orbiting objects are 14.6 $\mjup$ (Object1) and 30.1 $\mjup$ (Object2) respectively.  \label{fig:stable}}
\end{center}
\end{figure*}

An example of a configuration that remains stable can be seen in Figure \ref{fig:stable}.  This system consists of a 1.18 $\msol$ star, and 2 orbiting objects with masses of 14.6 $\mjup$ and 30.1 $\mjup$, and semimajor axes of 8.8 AU and 36.5 AU respectively.  This ordering of fragment masses - increasing mass with increasing separation - is a natural consequence of the Jeans mass formalism for initial fragment mass (see e.g. Figure 2 of \citealt{Forgan2011a}).  In this case, the mass ordering is preserved throughout the simulation, and both objects assume stable, low-eccentricity orbits (Figure \ref{fig:stable}).  This is primarily due to the more rapid orbital migration of the inner object separating the fragments by nearly 28 AU before self-interactions are activated.

\begin{figure*}
\begin{center}$
\begin{array}{cc}
\includegraphics[scale=0.4]{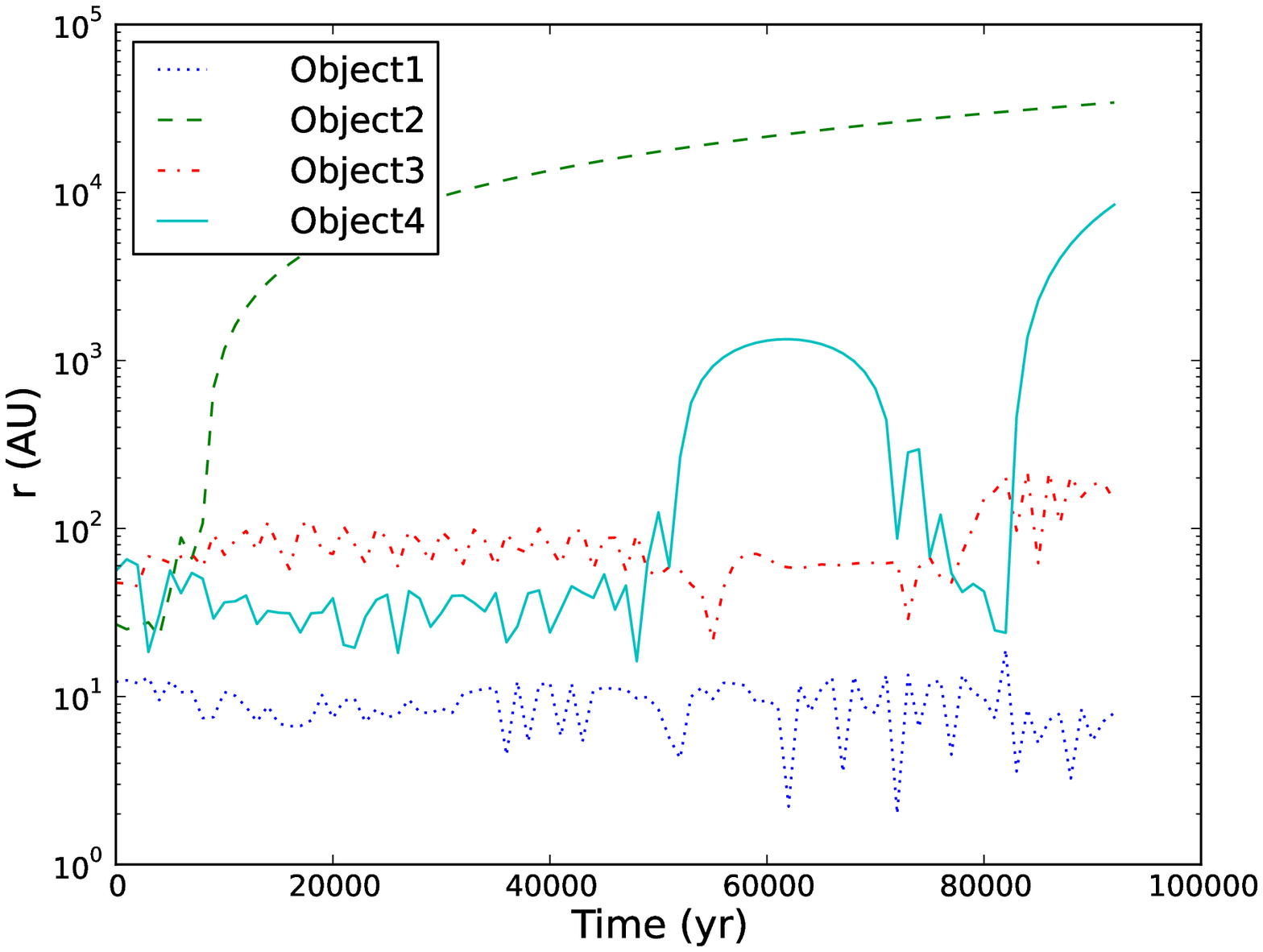} &
\includegraphics[scale=0.4]{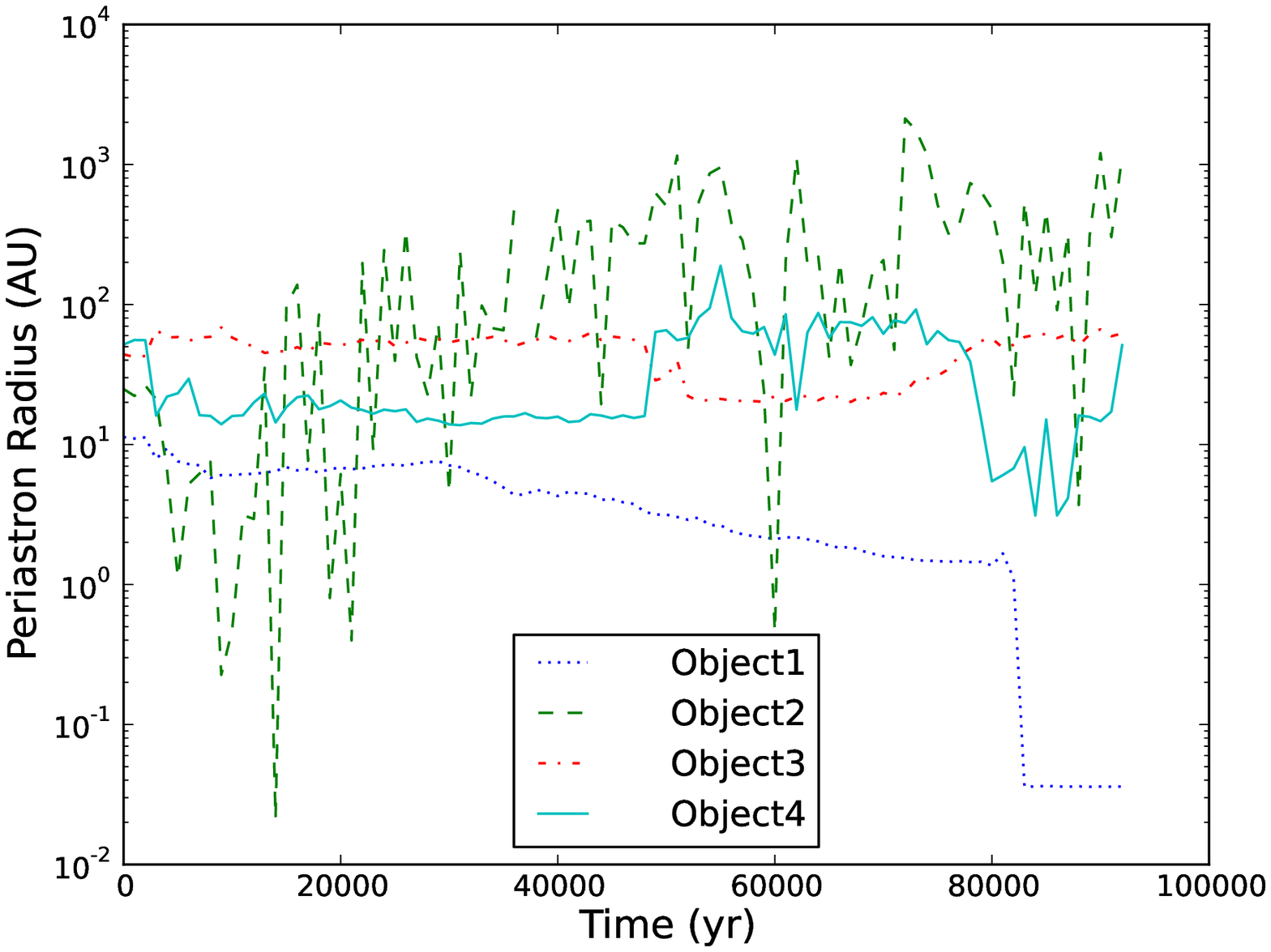} \\
\end{array}$
\caption{Distance from the star (left) and periastron radius (right) as a function of time for a single system that undergoes strong fragment-fragment scattering.  The central star is 0.8 $\msol$, with the 4 orbiting objects having masses 4.7 $\mjup$ (Object1), 2.8 $\mjup$ (Object2), 18.5 $\mjup$ (Object3) and 7 $\mjup$ (Object4). \label{fig:scattering}}
\end{center}
\end{figure*}

Figure \ref{fig:scattering} shows a system with a 0.8 $\msol$ star and 4 orbiting objects, with masses 4.7 $\mjup$, 2.8 $\mjup$, 18.5 $\mjup$ and 7 $\mjup$ and semimajor axes of 12.2, 26.9, 47.6 and 55.9 AU respectively.  The tidal downsizing process does not preserve the mass ordering of fragments, resulting in an unstable configuration.  After several orbit crossings in the first few thousand years, the object second-closest to the star is ejected, and the eccentricities of the remaining objects are boosted.  In this relatively extreme case, the periastron radius of object number 3 is reduced to of order a solar radius, and would almost certainly undergo Roche lobe overflow and be completely destroyed and accreted.  Object 1 also comes within a relatively close distance, and would likely have its eccentricity damped by tidal evolution (as well as undergoing significant radiative heating and evaporation).

\begin{figure*}
\begin{center}$
\begin{array}{cc}
\includegraphics[scale=0.4]{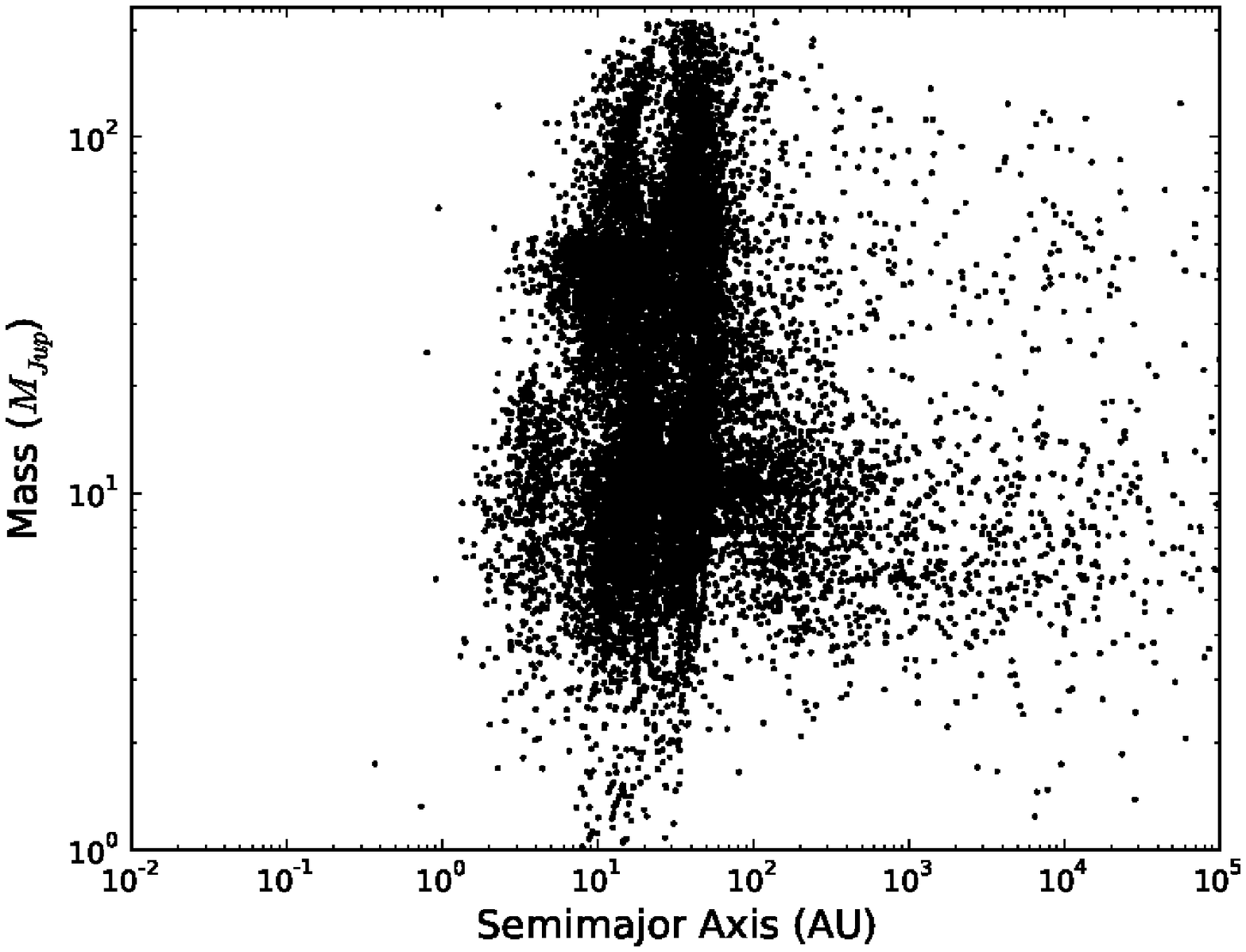} &
\includegraphics[scale=0.4]{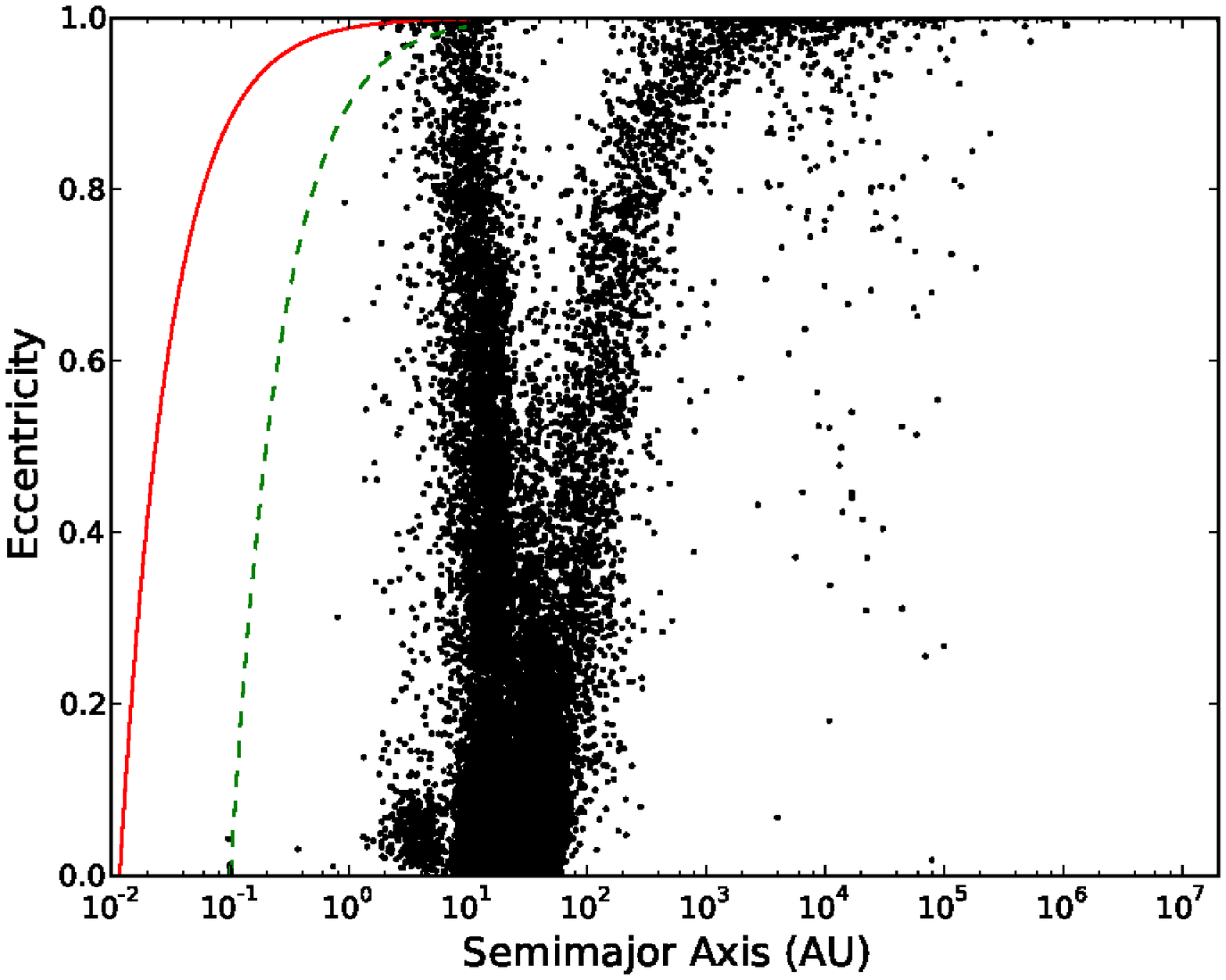} \\
\end{array}$
\caption{Mass versus semimajor axis after $10^6$ years of $N$-Body integration (left), and eccentricity versus semi-major axis after $N$-Body integration (right).  The initial eccentricity of all objects is zero.   \label{fig:mvsa_evsa}}
\end{center}
\end{figure*}

\begin{figure}
\begin{center}
\includegraphics[scale=0.4]{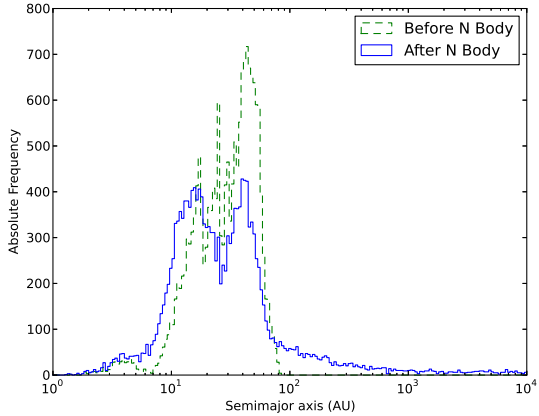}
\caption{Semimajor axis distribution for all fragments that remain bound to their host star before (dashed green line) and after (solid blue line) $10^6$ years of $N$-Body integration. \label{fig:ahist}}
\end{center}
\end{figure}

How are these processes encoded in the statistics of these bound objects? Figure \ref{fig:mvsa_evsa} shows the masses and eccentricities of the bound objects against semi-major axis at the end of the integration.  While the original distribution of objects with semimajor axis is largely preserved, with the median before and after integration being 29.9 and 24.2 AU respectively (see also Figure \ref{fig:ahist}), it is also apparent that scattering produces a long tail in the semimajor axis distribution out to $\sim 1 pc$, which skews the mean from 31.6 AU before integration to 2338.5 AU afterwards.

The eccentricity of all objects is initially zero, as the population synthesis model is not equipped to evolve this quantity.  After integration, the objects occupy the entire range of eccentricities permitting bound orbits ($0 \leq e<1$).  The two bands in the right panel of Figure \ref{fig:mvsa_evsa} are indicative of the typical spacing between two fragments in a 3-body system, the most common outcome of the population synthesis model.  The dashed green line indicates a periastron radius of 0.1 AU - objects which fall to the left of this line can be expected to undergo tidal interactions to reduce their eccentricity (although we do not model this effect).  The red line indicates a periastron radius of $1R_{\odot}$ - on some occasions, objects achieve periastra close to this value, and are likely to undergo Roche lobe overflow and be partially or completely disrupted.

\subsubsection{Properties of Ejected Objects}

\begin{figure}
\begin{center}
\includegraphics[scale=0.4]{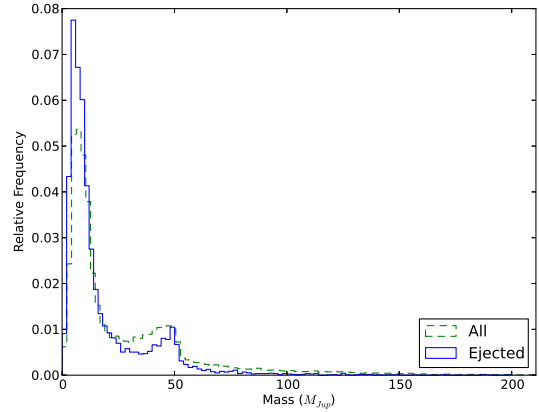}
\caption{Mass distribution of all fragments (dashed green line) and all ejected objects (solid blue line) after $10^6$ years of $N$-Body integration. \label{fig:mhist}}
\end{center}
\end{figure}

\noindent Of the 23,325 objects participating in these integrations, 5,956 achieve eccentricities greater than unity, and are ejected from the system.  Figure \ref{fig:mhist} compares the mass functions of all objects in the simulation (dashed green) to the objects ejected from their host systems (solid blue).  The general shape of both functions is similar - they share two peaks at $\sim 3 \mjup$ and $\sim 50 \mjup$.  However, it is clear that lower mass objects are preferentially ejected, enhancing the first peak and reducing the relative number of objects with masses greater than $\sim 20 \mjup$.  The origin of the two peaks is linked to the properties of the disc and the spacing between disc fragments.  As we have used G stars exclusively in this analysis, we only sample a small part of the mass function for disc fragment ejecta.  Increasing the mass of the star increase the typical fragment mass, however the stellar IMF will reduce their contribution to the final function.

\begin{figure*}
\begin{center}$
\begin{array}{cc}
\includegraphics[scale=0.4]{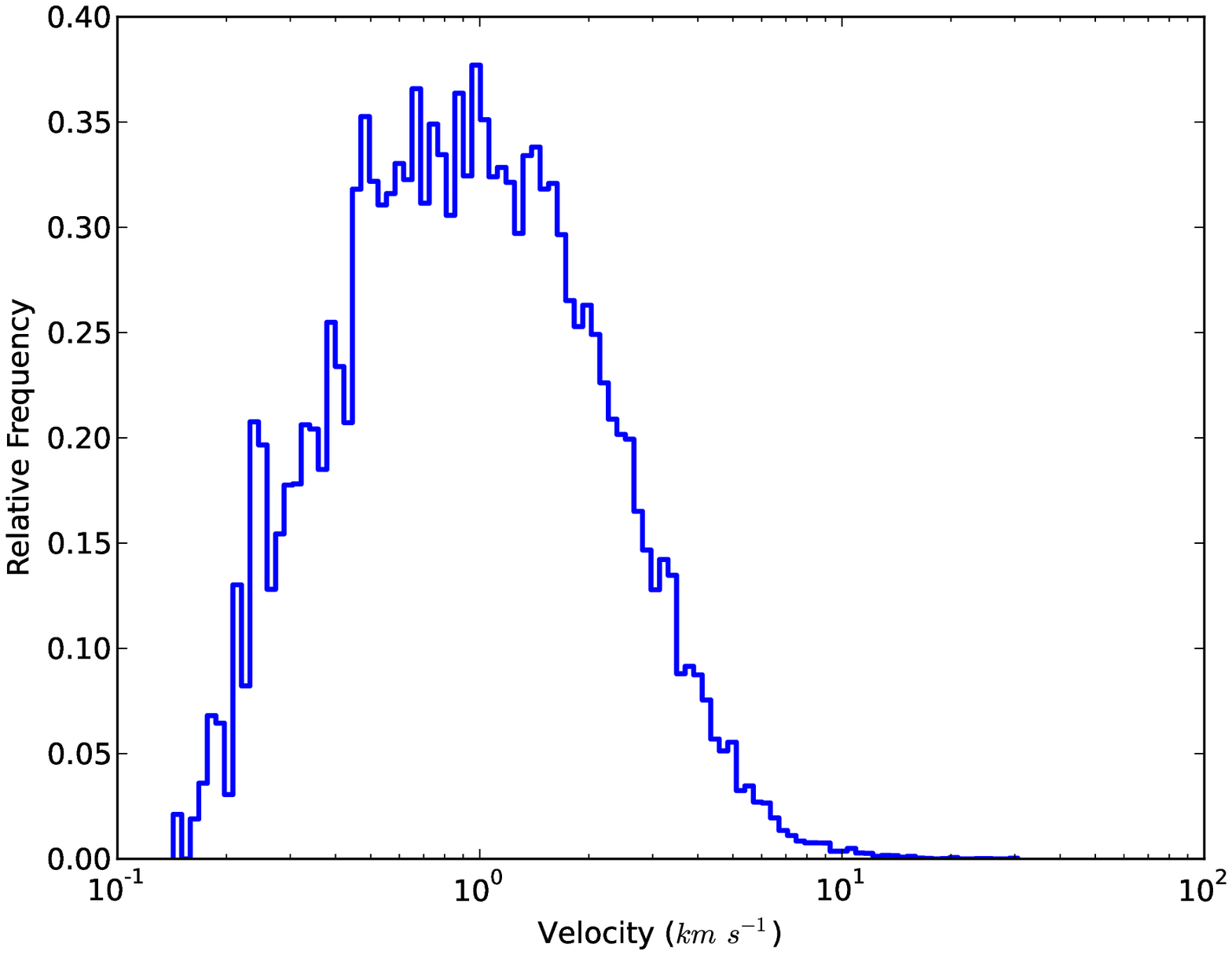} &
\includegraphics[scale=0.4]{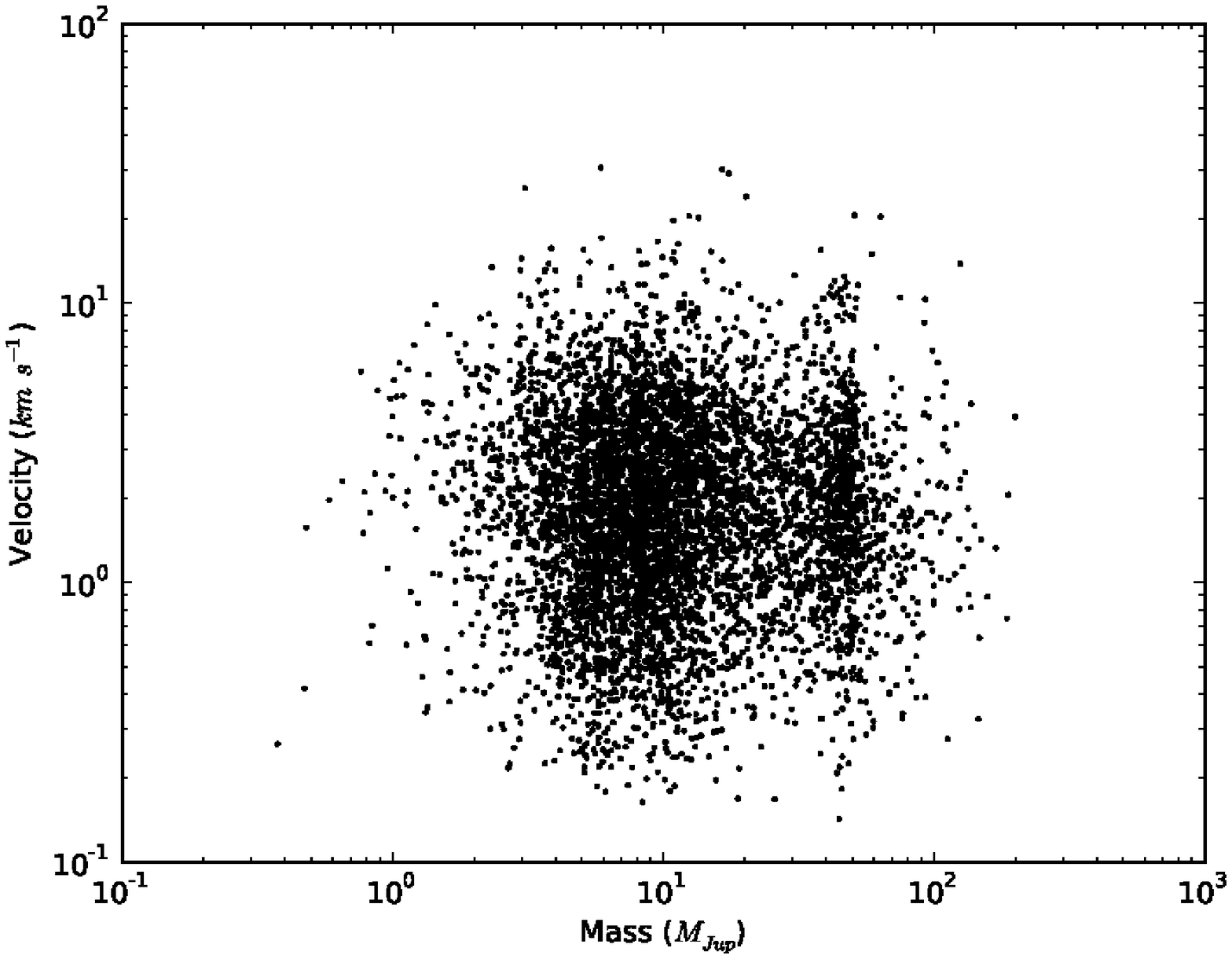} \\
\end{array}$
\caption{The velocity distribution of ejected objects (left), and the mass-velocity distribution of ejecta (right). \label{fig:ejectedmvsv}}
\end{center}
\end{figure*}

Figure \ref{fig:ejectedmvsv} shows the distribution of ejecta velocities, as well as the relationship between mass and velocity.  The mean velocity of the ejecta is 2.4 $km s^{-1}$, with a median of 1.85 $km s^{-1}$.  There appears to be no correlation between the ejecta mass and their velocity.  This is in accord with ensembles of 3D hydrodynamical simulations of disc fragmentation \citep{Rice2003a,Stamatellos2009}.

As we will see in the next section, these velocities are similar to those found in the cluster experiments, and as such we cannot diagnose the formation mechanism of a free floating body from velocity alone.

\subsection{The Effect of Cluster Environment}

\begin{figure}
\begin{center}
\rotatebox{270}{\includegraphics[scale=0.4]{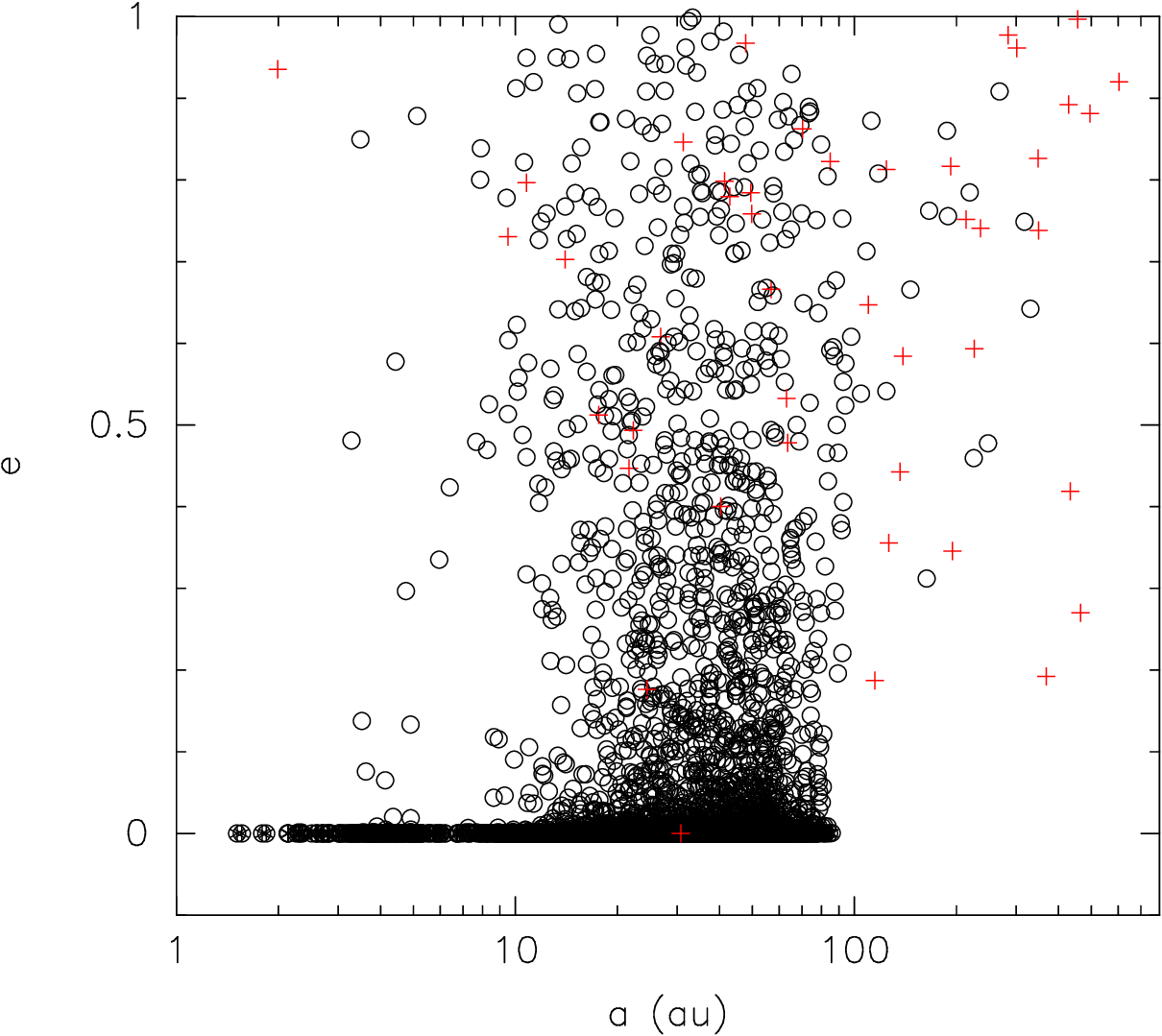}}
\end{center}
\caption[bf]{Orbital eccentricity versus semimajor axis for planets after 10\,Myr of dynamical evolution in star-forming regions. We have summed together 
the results of 10 simulations with the same initial conditions. Circles indicate `primordial' planetary systems
and crosses show captured planets.}
\label{ecc_sma_regions}
\end{figure}

We begin by placing the population from \citet{TD_synthesis} into our subvirial (collapsing) star-forming regions. In Fig.~\ref{ecc_sma_regions} we show the orbitial eccentricity versus semimajor axis after 10\,Myr of evolution. The open circles are primordial, or birth systems -- the planet is still orbiting its parent star. The red plus signs represent captured systems -- the planets first become free-floating and then are captured into an orbit around a low- or intermediate-mass star ($m < 2.5$M$_\odot$) which no longer has its planet, or a more massive star which has lost its stellar partner and has accquired a planet. As in \citet{Parker2012}, the captured planets typically have high eccentricity and high inclinations (often retrograde orbits) and their semimajor axes are often 100s -- 1000s\,au \citep[see also][]{Perets2012}.

The fraction of planets that become free-floating is $f_{\rm FFLOP} = 0.13 \pm 0.03$, the fraction that have their eccentricity raised to $e > 0.1$ is $f_{\rm e, alt} = 0.12 \pm 0.02$ and the fraction that have their semimajor axis altered by $\pm$10\,\% is $f_{\rm a, alt} = 0.057 \pm 0.009$. These fractions are not too dissimilar to those reported in \citet{Parker2012} for planets originally on 30\,au orbits, and the median semimajor axis in the distribution from \citep{TD_synthesis} is of order 30\,au.

\begin{figure*}
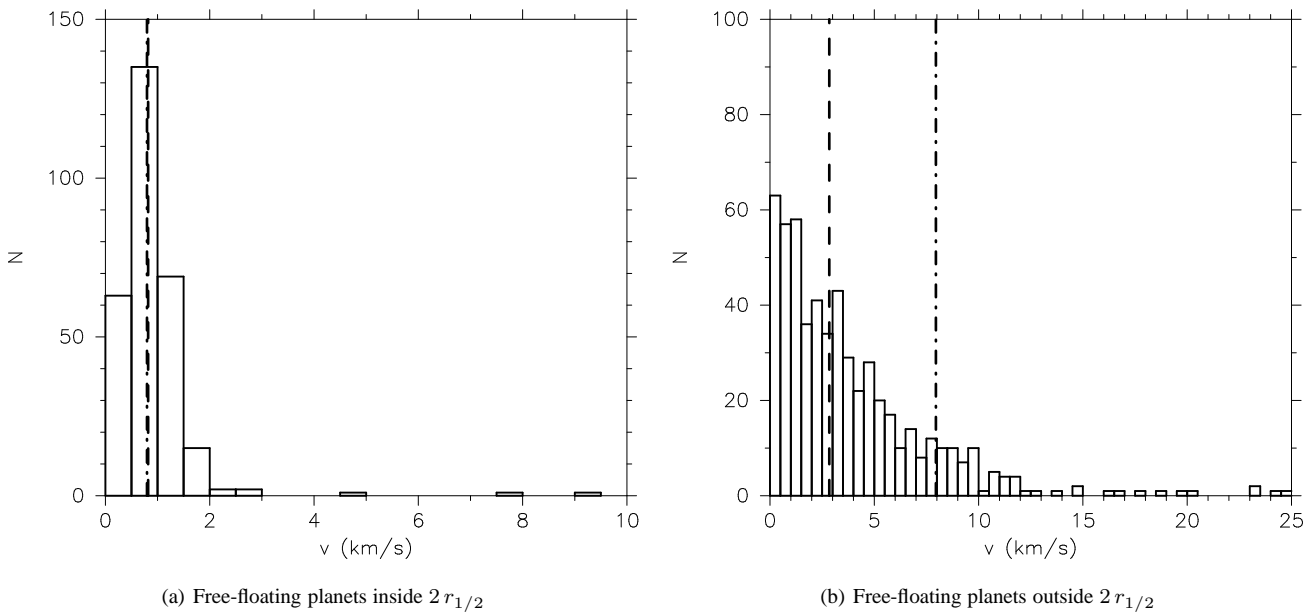

  \begin{center}
\setlength{\subfigcapskip}{10pt}
\hspace*{-1.5cm}\subfigure[Free-floating planets inside $2\,r_{1/2}$]{\rotatebox{270}{\includegraphics[scale=0.4]{Plot_Or_SP_C_F2p0_S1FF10_vel_bound.ps}}}
\hspace*{0.3cm} 
\subfigure[Free-floating planets outside $2\,r_{1/2}$]{\rotatebox{270}{\includegraphics[scale=0.4]{Plot_Or_SP_C_F2p0_S1FF10_vel_esc.ps}}} 
\caption[bf]{The distribution of velocities of free-floating planets
  that are outside two half-mass radii from the centre of the cluster
  ($2\,r_{1/2}$) after 10\,Myr. The median 3-D velocity for planets is shown by the dashed
  line whereas the median 3-D velocity for stars AND planets shown by dot-dashed line.}
\label{vel_regions}
  \end{center}
\end{figure*}

We show the velocity distribution of the free-floating planets after 10\,Myr in Fig.~\ref{vel_regions}.  Panel a) gives the 3D velocity magnitude for planets still `observationally bound' to the cluster (within two half-mass radii, $2r_{1/2}$ of the centre), and panel b) gives the same property for the objects that are unbound (outside of two half-mass radii of the centre) in panel (b). We show the median velocity of the planets by the dashed line, and the median velocity of both stars and planets in the region by the dot-dashed line. 

The free-floating planets that are still bound to the cluster have very similar velocities to the median, whereas those that are ejected from the cluster have much lower velocities than stars that are also ejected from the cluster. Again, this is similar to the result reported in \citet{Parker2012} for planets originally on 30\,au orbits. 

We have seen that many of the planets from the population synthesis in \citet{TD_synthesis} are ejected onto wide orbits through dynamical interactions in star-forming regions that subsequently form a star cluster. As discussed in Section~\ref{sec:method:cluster}, fragment--fragment scattering acts to increase the eccentricity and semimajor axis of the planets at the same time as the dynamical evolution of the star-forming regions. For this reason, it is difficult to disentangle the effects of fragment--fragment scattering from interactions in the star-forming region. 

\begin{figure}
\begin{center}
\rotatebox{270}{\includegraphics[scale=0.4]{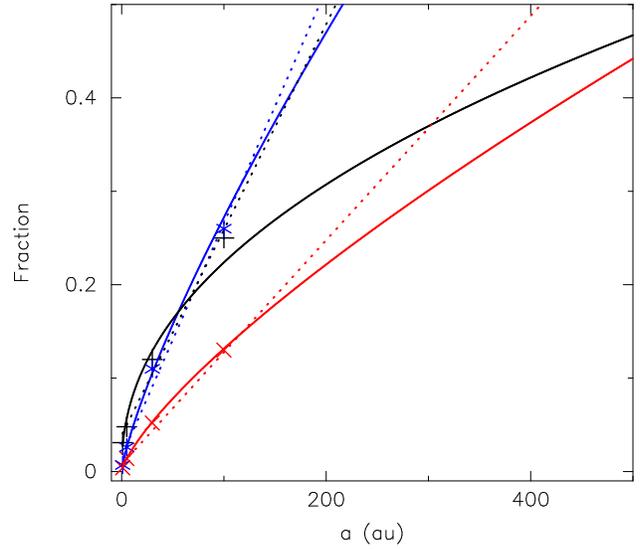}}
\end{center}
\caption[bf]{Fraction of systems after 10\,Myr of dynamical evolution in star-forming regions that are free floating (the black plus signs), that have their eccentricity altered to above 0.1 (the blue asterisks) and those that have their semimajor axis altered by more than 10\,per cent (the red crosses), for planets initially at 1, 5, 30 and 100\,au from the host star. We also show fits to the points via the coloured lines; solid lines are power law fits, and dotted lines are linear fits.}
\label{fiducial_regions}
\end{figure}

The probability of dynamical interactions in a star-forming region removing a free-floating planet \emph{after} fragment--fragment scattering onto a wider orbit can be approximated from the result of simulations in which we place single planets at 1, 5, 30 and 100\,au. We show the fraction of free-floating planets (the black plus signs), and the fractions of planets with significantly altered semimajor axes (red crosses) and eccentricities (blue asterisks), as a function of semimajor axis is shown in Fig.~\ref{fiducial_regions}.

The fraction of free-floating planets as a function of initial semimajor axis, $a$, can be approximated by a linear relation (the dotted black line), but a better fit is a power law where  

\begin{equation} 
f_{\rm FFLOP} = 0.027a^{0.46},
\end{equation} 

\noindent meaning that 50\% of planets would be free-floating if their semimajor axis were greater than 570\,au, and all planets with $a > 8350$\,au would become free-floating in these dense regions.  Similarly, 

\begin{equation}
f_{\rm e, alt} = 0.0072a^{0.79},
\end{equation}

meaning 50\% of planets would have $e>0.1$ if their semimajor axis were greater than 214\,au and,

\begin{equation}
f_{\rm a, alt} = 0.0042a^{0.75},
\end{equation} 

\noindent and as such 50\,per cent of planets would have their semimajor axis altered by $a \pm $10\,per cent if their semimajor axis were initially 586\,au.

We note that the above fractions pertain only to planets formed from tidal downsizing, and do not include planets which may form on much closer (and hence more tightly bound) orbits.

It is unlikely that self-gravitating discs could maintain outer radii of several hundreds of AU due to truncation by stellar encounters \citep{Rosotti2014}, but the scattering simulations indicate that fragments can be boosted to orbits where the probability of becoming free-floating is significant. The mean semimajor axis of fragments still bound to the host after 1 Myr of fragment-fragment scattering is significantly larger than the 50\% threshold values quoted above (albeit the median remains much lower).  

Finally, we note that the mass of the planet is unimportant when determining the fraction of systems that are liberated form their host star in a clustered environment. The reason for this is that the planetary mass is almost negligible with respect to the host star, and any impulsive encounter that removes the planet is usually orders of magnitude more energetic than the mass-dependent binding energy of the system \citep{Fregeau2006,Parker2013b}.

\section{Discussion} \label{sec:discussion}

\subsection{Limitations of the Analysis}

\subsubsection{Fragment-Fragment Scattering}

\noindent Perhaps the most important outstanding issue with the isolated fragment-fragment scattering simulations is that fragment interactions are switched on after the disc is dissipated.  The population synthesis model makes a simple approximation of a single fragment's migration through the disc, thanks to the torques produced by the spiral arms that the fragment excites.  Multiple fragments will excite multiple spiral features, and hence the fragments' migration will depend on the morphology and evolution of these features.  We therefore cannot fully capture evolution such as resonant migration while the disc is present, which could increase the probability of mean motion resonances in these isolated systems.

We assume that no further evolution of the fragment's internal structure occurs during the $N$-Body integrations.  The grains trapped within the envelopes of the fragments may continue to grow and evolve long after the disc has dissipated.  For those objects which have internal temperatures surpassing the grain sublimation temperature, this may be a reasonable approximation.  However, it is clear that fragments undergoing close encounters may experience tidal forces strong enough to disrupt equilibrium density profiles.  In the extreme case, fragments that collide and merge may produce more massive objects, with the subsequent shocks affecting the object's chemistry and thermodynamic balance.  We do not model collisions in this work, and so this behaviour remains uninvestigated. 

Even discounting collisions, close encounters between two fragments may also affect their geochemistry.  The integrations do not include the effects of tides being raised on the surface of the fragments, which may damp the fragment's eccentricity, and dissipate heat in its interior.  Close encounters will produce significant tides, and will likely cause significant melting or sintering of any solid core inside a fragment, which will have consequences for observed terrestrial planets if the envelope is stripped (cf \citealt{Nayakshin2014}).


\subsubsection{Cluster Simulations}

Our simulations of cluster environments and fragment evolution have several limiting factors. Firstly, the stars (and planets) are assumed to form instantaneously, which is obviously unrealistic. However, hydrodynamical calculations of star-formation are still limited by computing power, meaning that a large statistical sample of simulations cannot be achieved. They also typically only follow the first $\sim$Myr of star formation \citep[e.g.][]{Bate2012} and so cannot be evolved past the point at which dynamical interactions become negligible.

Ideally, we would also like to follow the effects of multi--fragment scattering \emph{and} the effects of stellar fly-bys simultaneously. Efforts have been made to implement this in full cluster simulations \citep[e.g.][]{Hao2013,Pacucci2013,Liu2013}, but in the meantime we can learn much from single-fragment evolution and also use collisional cross sections to make analytical estimates for the effect of the cluster environment on planetary systems \citep[e.g.][and many more]{Laughlin1998,Bonnell2001,Adams2006,Parker2012,Craig2013}.

Interestingly, the fraction of extrasolar planetary systems that could have been affected by dynamical interactions in the natal star-forming environment is almost totally unconstrained. In the local solar neighbourhood, \citet{Bressert2010} used the local surface density around YSOs to argue that most nearby star-forming regions are low-density $<$100\,stars\,pc$^{-2}$ and protoplanetary discs and systems are likely to be unaffected by dynamical evolution. However, it is unclear whether the \citet{Bressert2010} sample is representative of most star-forming events that produced the Galactic field population and even if it is, \citet{Parker2012d} argue that 50\,per cent of local star-forming regions could be dense enough to affect planetary systems.

The simulated star-forming regions presented here typically have maximum median densities of $\sim 5000$\,stars\,pc$^{-2}$, and we suggest that our simulations are probably an upper limit on the amount of dynamical evolution that may have affected young planetary systems.

\subsection{Implications for Observations}

\noindent Despite a great deal of dynamical evolution, the majority of objects formed through disc instability exist at large semimajor axes.  Even if the objects form with low eccentricity, this can be pumped up to high eccentricity through fragment-fragment scattering and dynamical interactions with the parent cluster.  This implies that current exoplanet detection missions remain relatively insensitive to disc instability objects, with the exception of microlensing and direct imaging surveys \citep[eg][]{Bonavita2014}.  While planets captured onto high inclination orbits are not exclusively disc fragments - for example, the Lidov-Kozai mechanism can produce high inclination orbits for a variety of planet parameters \citep{Naoz2013} -  detecting giant planets and brown dwarfs with large semimajor axes, eccentricities and inclinations are supporting evidence for their early formation while inside their parent cluster, a key attribute of gravitational instability.

Equally, the observed ejection rate of objects due to scattering, and the efficacy of stripping by a cluster for semimajor axes above $\sim 100$ AU, confirms that a large fraction of disc instability objects will exist as free-floating planets and field brown dwarfs.  The velocity dispersion of these objects will be relatively low compared to that of the parent cluster (at least initially).  These targets are also amenable to microlensing surveys, as well as infrared surveys of substellar objects \citep[eg][]{Best2013,Pinfield2013,Beichman2014}.  

\section{Conclusions }\label{sec:conclusions}

\noindent We have investigated the dynamical fate of self-gravitating disc fragments once the disc has dissipated.  By carrying out separate $N$-Body integrations of a) isolated systems undergoing fragment-fragment scattering and b) systems embedded in substructured star-forming regions (which collapse to form clusters) we show that the descendants of disc fragments fill a large area of parameter space associated with the orbital elements.  While most objects retain semimajor axes and eccentricities that are similar to those of their formation, a significant fraction experience strong orbit modification, in some cases being ejected from their parent systems.  Some are recaptured on highly inclined, eccentric orbits.  Others are scattered inward toward the parent star, and are likely to undergo Roche lobe overflow and be destroyed.

By combining these calculations with the results of our population synthesis models \citep{TD_synthesis}, we can see that disc fragments share a variety of fates.  Over a third are completely destroyed by tidal disruption (see also \citealt{Zhu2012}). Of the surviving fragments, at least a quarter are ejected from the system - if fragment-fragment scattering and cluster dynamics were considered in the same integration, it is likely this figure would be higher.  The remainder tend to occupy orbits with semimajor axes greater than 10 AU, with eccentricities spanning all possible values for bound orbits.  Any planets that are to subsequently form in these  systems via core accretion must do so in the shadow of this dynamical evolution, and under the perturbative forces generated by the descendants of disc fragments.  

Characterising the mass and velocity distributions of the free-floating planet population, especially in the 1-10 $\mjup$ range where tidal downsizing is favoured and core accretion disfavoured, will provide important constraints on the frequency of fragmenting systems, to complement direct imaging surveys for massive objects still bound to their parent star (e.g. \citealt{Bonavita2014, Bowler2014, Desidera2014}).  This work presents the first statistically significant sample of disc fragment properties post-ejection for observers to help estimate this frequency.  Any frequency derived from observations will depend sensitively on both the ejection rate and fragment destruction rate.  Further work must refine estimates for both of these properties.

\section*{Acknowledgments}

DF and KR acknowledge support from STFC grant ST/J001422/1.  DF also acknowledges support from the ECOGAL ERC Advanced Grant Programme.  RJP acknowledges support from the Royal Astronomical Society in the form of a research fellowship.

\bibliographystyle{mn2e} 
\bibliography{TD_dynamics}

\appendix

\label{lastpage}

\end{document}